\documentclass[fleqn,usenatbib]{mnras}

\usepackage[T1]{fontenc}
\usepackage{xcolor}
\DeclareRobustCommand{\VAN}[3]{#2}
\let\VANthebibliography\thebibliography
\def\thebibliography{\DeclareRobustCommand{\VAN}[3]{##3}\VANthebibliography}

\usepackage{graphicx}	% Including figure files
\usepackage{amsmath}	% Advanced maths commands
\usepackage{amssymb}	% Extra maths symbols
\usepackage{lineno}
\usepackage[normalem]{ulem}

\newcommand{\gaia}{\textit{Gaia}}
\newcommand{\kms}{km~s$^{-1}$}
\newcommand{\kmskpc}{km~s$^{-1}$~kpc$^{-1}$}
\newcommand{\HI}{H\,\textsc{i}}

\newcommand{\edit}[2]{#2}
\newcommand{\editt}[2]{#2}
\newcommand{\edittt}[2]{#2}

\defcitealias{ramos18}{R18}
\defcitealias{lucchini23}{Paper I}

\title[The Milky Way Bar with Hercules]{\edittt{Constraining }{}The Milky Way Bar \edit{Length}{Pattern Speed} using Hercules and \textit{Gaia} DR3}

\author[Lucchini, et al.]{
	Scott Lucchini$^{1,2\dagger}$, Elena D'Onghia$^{1,3}$, J. Alfonso L. Aguerri$^{4,5}$ 
	\\
	$^{1}$Department of Physics, University of Wisconsin - Madison, Madison, WI, USA \\
        $^{2}$Center for Astrophysics | Harvard \& Smithsonian, Cambridge, MA, USA \\
	$^{3}$Department of Astronomy, University of Wisconsin - Madison, Madison, WI, USA \\
	$^{4}$Instituto de Astrofísica de Canarias; C/ Vía Láctea s/n, 38200, La Laguna, Spain\\
    $^{5}$Departamento de Astrofísica, Universidad de La Laguna, E-38206 La Laguna, Spain \\
    $^\dagger$scott.lucchini@cfa.harvard.edu
}

\date{Accepted 2024 March 26. Received 2024 March 18; in original form 2023 May 5}

\pubyear{2023}

\usepackage{newtxtext,newtxmath}
\usepackage{cancel}
\begin{document}
	\label{firstpage}
	\pagerange{\pageref{firstpage}--\pageref{lastpage}}
	\maketitle

\begin{abstract}
The distribution of moving groups in the solar neighborhood has been used to constrain dynamical properties of the Milky Way for decades. \editt{Unfortunately, the unique}{The kinematic} bimodality between the main mode (Hyades, Pleiades, Coma Berenices, and Sirius) and Hercules can be explained by two different bar models -- via the outer Lindblad resonance of a \editt{short, fast }{}bar \editt{}{with a high pattern speed ($\sim$55 \kmskpc)}, or via the corotation resonance of a \editt{long, slow }{}bar \editt{}{with a low pattern speed ($\sim$40 \kmskpc)}. \editt{While various recent works have used alternative methods to converge on the slow bar model, in this work, we explicitly break this degeneracy by}{Recent works directly studying the kinematics of bar stars and gas flows near the center of the Galaxy have converged on the low pattern speed model. In this paper, we independently confirm this result by} using \gaia\ DR3 to \editt{explore}{directly study} the variation of Hercules across Galactic azimuth. We find that Hercules increases in $V_\phi$ and becomes stronger as we move towards the minor axis of the bar, and decreases in $V_\phi$ and becomes weaker as we move towards the major axis of the bar. This is in direct agreement with theoretical predictions of a \editt{long, slow}{low pattern speed} bar model in which Hercules is formed by the corotation resonance with stars orbiting the bar's L4/L5 Lagrange points.
\end{abstract}


% \linenumbers

\section{Introduction} \label{sec:intro}

Being embedded within the Milky Way (MW) makes it difficult to observe its properties. We must use indirect methods to determine the distribution of mass, and corresponding nonaxisymmetric features in our Galaxy. Specifically, \editt{}{determining} the properties of our own Galactic bar -- an oblong stellar overdensity at the center of our Galaxy -- \editt{are still not known for sure}{have required several independent techniques to converge on a consensus}.
\editt{There have been two main models proposed throughout the past decades, one in which we have a short bar that rotates very quickly ($R_\mathrm{bar}\sim3$ kpc, $\Omega_p\sim55$ \kmskpc; \mbox{\citealt{dehnen00,debattista02,monari17}}; \mbox{\citealt{fragkoudi19}}), and one in which we have a long bar that rotates relatively slowly ($R_\mathrm{bar}\sim5$ kpc, $\Omega_p\sim40$ \kmskpc; \mbox{\citealt{perez-villegas17,monari19,asano20}}; \mbox{\citealt{donghia20}}).}{Three main methods that have been used are: direct observations of the kinematics of bar stars, gas flows in the inner Galaxy, and the kinematics of stars in the solar neighborhood (i.e. ``moving groups'').}

\editt{Originally, the short, fast bar scenario proposed by \mbox{\citet{dehnen00}} was supported by an application of the Tremaine-Weinberg method \mbox{\citep{tremaine84}} modified for use with radial velocities in which \mbox{\citet{debattista02}} used OH/IR stars to measure a pattern speed of $\Omega_p\sim60$ \kmskpc. The gas motions measured in \HI\ Galactic longitude vs velocity ($\ell v$-diagram) in the inner Galaxy were also consistent with a fast bar scenario \mbox{\citep{englmaier97,fux99}}. However, star counts have indicated that the bar may be long, extending near or past the proposed corotation resonance of a fast bar \mbox{\citep{benjamin05,wegg15,clarke19}}. Moreover, recent studies using more accurate velocity data have found lower values of $\Omega_p\sim40$ \kmskpc consistent with a longer bar \mbox{\citep{sanders19,bovy19}}. A more thorough parameter-space exploration using 2D isothermal simulations in external potentials have also shown that slow bars actually provide the best match to all the $\ell v$-diagram features \mbox{\citep{sormani15}}.}{}

\editt{}{One of the first attempts to directly measure the bar pattern speed used a version of the Tremaine-Weinberg method \citep{tremaine84} modified for use with radial velocities \citep{debattista02}. Using OH/IR stars, they found a high pattern speed of $\Omega_p\sim60$ \kmskpc. However recently, more advanced MW surveys have allowed us to directly observe the transverse kinematics of stars towards the center of our Galaxy \citep{portail17,sanders19,clarke22}, and these studies all support at low pattern speed of $\Omega_p\sim30-40$~\kmskpc.}

\editt{}{Similarly, gas kinematics towards the Galactic center were also initially interpreted as a signature of a high $\Omega_p$ bar \citep{fux99,bissantz03}. Comparing the observed Galactic longitude vs \HI\ radial velocity diagrams ($\ell v$-diagrams) against simulations, \citet{bissantz03} found a best fit with $\Omega_p=60\pm5$~\kmskpc. However, improved simulation techniques have shown that again, a low pattern speed is more consistent with the data ($\Omega_p\sim30-40$~\kmskpc; \citealt{sormani15,li16,li22}).}

\editt{An additional constraint on the properties of the MW's nonaxisymmetric features came from the motions (in full 3D) of the stars near the Sun \mbox{\citep[e.g.][]{dehnen98,antoja12}}.}{Using the full 3D velocities of stars near the Sun, we can also investigate the inhomogeneities in the velocity distributions to learn about the nonaxisymmetric features of our Galaxy.}
How these stars cluster in velocity space (moving groups) contains a wealth of information about the evolution of our Galaxy's disk and the forces these stars are feeling. Specifically, the bimodality in the Galactocentric azimuthal velocity versus radial velocity plot has been one of the main features models have attempted to reproduce. The main mode (at $V_\phi\sim-230$~\kms) contains the moving groups of Hyades, Pleiades, Coma Berenices, and Sirius, while Hercules (at $V_\phi\sim-200$~\kms) is separated by a gap (a strong underdensity).

This bimodality has been explained through resonances of the MW's bar. Since the bar contributes a nonaxisymmetric gravitational potential, it has an associated pattern speed, as discussed above. 
Over time, the bar perturbs the stars to align the frequencies of the stellar orbits with the bar's frequency. These stars are ``in resonance'' with the bar. A bar's strongest resonances are the corotation resonance (CR; in which the stellar and bar frequencies match exactly), and the inner and outer Lindblad resonances (ILR and OLR; in which the star completes two radial oscillations for every orbit around the galaxy).
Unfortunately, \editt{both proposed models of the Galactic bar can produce this bimodality through different resonances}{depending on the pattern speed of the bar, these resonances fall at different locations, and multiple models are able to explain this bimodality}. For a \editt{short bar}{high pattern speed bar ($\Omega_p\sim55$~\kmskpc)}, the OLR falls just inside the solar neighborhood \edittt{and is able to trap stars to form Hercules}{with Hercules being formed by stars moving outward in the disk coming from inside $R_\mathrm{OLR}$} \citep{dehnen00,minchev10,antoja14,fragkoudi19}. For a \editt{long bar}{low pattern speed bar ($\Omega_p\sim40$~\kmskpc)}, the CR is able to create Hercules through stars orbiting the bar's Lagrange points \citep{perez-villegas17,hunt18,monari19,asano20,donghia20}. \editt{Both models agree with the observations near the Sun, however as we move throughout the Galactic disk their predictions change.}{While most modern models are more consistent with a low pattern speed, a direct observation using stellar kinematics indicating Hercules having been formed via CR or via OLR has remained elusive.}

\edit{}{Recent works explored various bar's resonances to discriminate between the models. \citet{monari19} analytically explored the impact of the $m=3$, 4, and 6 modes in addition to the $m=2$ CR and OLR modes. Finding that many of these other resonances aligned with ridges and structures in the local velocity plane, they determined that a \editt{slow}{low pattern speed} bar ($\Omega_p=39$~\kmskpc) provides the best fit. Furthermore, analysis of local gas flows and metallicity gradients also indicate that Hercules has been formed by \editt{the }{}CR \editt{of a long, slow bar }{}\citep{binney20,chiba21}. \edittt{}{An in-depth study of the effect of resonance trapping on Hercules also concludes that a slowing bar with a current pattern speed of 35~\kmskpc\ provides the best fit to the data \citep{chiba21b}.} Additionally, \citet{antoja14} was able to use RAVE data to explore the behavior of Hercules beyond the solar neighborhood varying radius. While they find that their data is more consistent with a \editt{fast}{high pattern speed} bar model ($\Omega_p=54$~\kmskpc), there is little difference in the R-dependence for the OLR and the CR.}

\editt{}{While both the CR and the OLR have similar signatures in the \edittt{}{$V_R-V_\phi$ plane in the} solar neighborhood, analytical calculations and numerical simulations have shown that the predictions of these two models differ as we move throughout the Galactic disk \citep{monari19herc,donghia20}.}
With \gaia's latest data release (Data Release 3; \citealt{gaia,gaia18,gaiadr3}), we can \editt{also}{} begin to differentiate between these two models by looking \edit{at full 3D stellar motions across the Galactic disk}{beyond the solar neighborhood in azimuth}. In this paper, we \edit{}{use \gaia's full 6D phase space information to} explore how Hercules changes as we move around the disk in \edit{azimuth}{$\phi$}, \editt{breaking the degeneracy between these two models for}{independently \edittt{constraining}{confirming} the pattern speed of} the MW's bar.

\section{Methods} \label{sec:methods}

\begin{figure*}
    \centering
    \includegraphics[width=\textwidth]{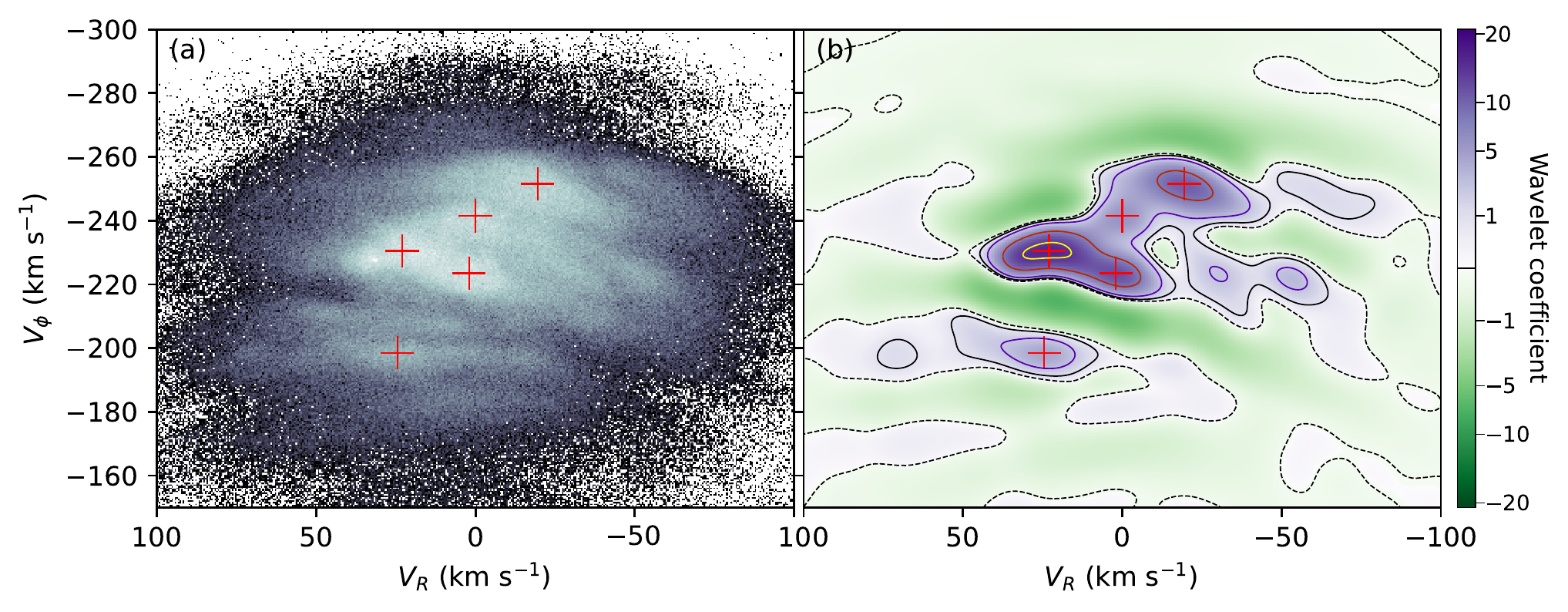}
    \caption{Panel a shows the 2D histogram of all the solar neighborhood stars in the $V_R-V_\phi$ kinematic plane using \gaia\ DR3 with a bin size of 0.5~\kms. Panel b shows the resultant wavelet transformed image at a scale of $8-16$~\kms. The purple and green regions depict the relative strength of the positive and negative wavelet coefficients, respectively. Contours are shown at the $-0.1$\% (dashed), 5\%, 15\%, 50\%, and 90\% levels. The plus signs denote the locations of the five major moving groups as detected by our wavelet transformation -- Hyades, Pleiades, Coma Berenices, Sirius, and Hercules.
    }
    \label{fig:results}
\end{figure*}

\begin{figure}
    \centering
    \includegraphics[width=\columnwidth]{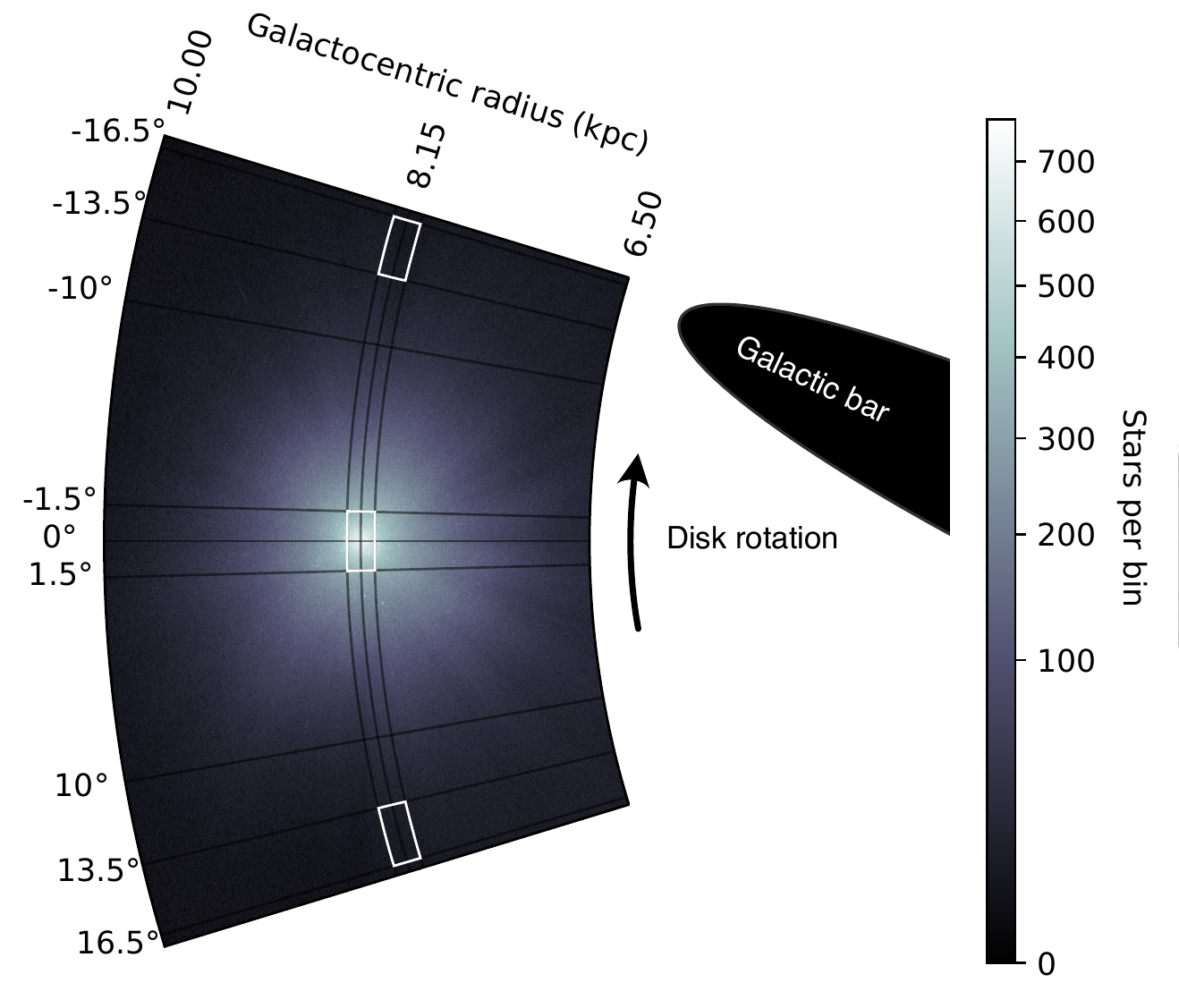}
    \caption{\gaia\ DR3 dataset in the context of the MW disk. This region extends from Galactocentric radius $R=6.5$~kpc to $R=10$~kpc, and from median azimuth $\phi=-15^\circ$ to $\phi=15^\circ$. An approximation of the Galactic bar is also shown. The MW disk is rotating \edit{counter-clockwise (in the direction of $+\phi$)}{clockwise (in the direction of $-\phi$)}, and the solar neighborhood is trailing behind the major axis of the bar. The white boxes also show the size of our neighborhood regions, extending 0.2 kpc in $R$ and $3^\circ$ in $\phi$. \edit{}{This coordinate system matches that of \citet{ramos18}.}}
    \label{fig:galaxyschematic}
\end{figure}

We follow the methods outlined in \citet{lucchini23} (hereafter \citetalias{lucchini23}) for \gaia\ selection and analysis using the wavelet transformation. Figure~\ref{fig:results} shows the solar neighborhood velocity plane and its corresponding wavelet transformed image. The plus signs denote the five most significant classical moving groups as detected by local maxima in the wavelet image. Starting with the 33,653,049 stars with radial velocities and geometric distances computed by \citet{bailer-jones21} \citep{bailer-jones-data}, we transformed the six-dimensional data into Galactocentric cylindrical coordinates\footnote{As in \citetalias{lucchini23}, we assume the Sun is located at \edit{$R_\odot=8.15$ kpc, $\phi_\odot = 0^\circ$, and $Z_\odot=5.5$ pc}{$R_0=8.15$ kpc, $\phi_0 = 0^\circ$, and $Z_0=5.5$ pc}, with velocity $(V_R, V_\phi, V_Z) = (10.6, -246.7, 7.6)$ \kms\ \edit{}{\citep{reid19}}.}. See Figure~\ref{fig:galaxyschematic} for a schematic defining our coordinate system with $R$ increasing away from the Galactic center, $\phi$ increasing \edit{in}{counter to} the direction of rotation, and $Z$ increasing towards the Galactic north pole. With the Sun located at \edit{$\phi_\odot=0^\circ$}{$\phi_0=0^\circ$}, this means the major axis of the Milky Way's bar is at \edit{$\phi\sim 20^\circ$}{$\phi\sim -20^\circ$} \citep{drimmel22}. Figure~\ref{fig:galaxyschematic} also shows the extent of the \gaia\ DR3 data that we used: $6.5 < R < 10$ kpc, and $-15^\circ < \phi < 15^\circ$. As in \citetalias{lucchini23}, we used the \textsc{pyia}\footnote{\url{https://github.com/adrn/pyia}} code to propagate the errors (including correlations) from the source properties (right ascension, declination, proper motions, and radial velocities) to the final properties (Galactocentric cylindrical coordinates; \citealt{pyia}).

For the analysis of Hercules in this paper, we broke up this region into 31 overlapping bins in $\phi$ of size \edit{$\phi=3^\circ$, $R=0.2$ kpc, and $Z=1$ kpc, centered on $R=8.15$ kpc and $Z=0$, spaced every $1^\circ$}{$\Delta\phi=3^\circ$, $\Delta R=0.2$ kpc, and $\Delta Z=1$ kpc, centered on $R=R_0=8.15$ kpc, $Z=0$, and various $\phi$ spaced by $1^\circ$} (e.g. ranges of $(-16.5^\circ,-13.5^\circ)$, $(-15.5^\circ,-12.5^\circ)$, etc).
% Figure~\ref{fig:starcounts} shows the number of stars in each of these bins as a function of azimuth.
The solar neighborhood bin contains nearly $10^6$ stars, while the bins at the edge of our sample ($\pm15^\circ$) contain more than $10^4$ stars.

We use the wavelet transform code, \textit{MGwave}\footnote{\url{https://github.com/DOnghiaGroup/MGwave}} (described in \citetalias{lucchini23}), to analyze the velocity distributions of these different neighborhoods. \edit{}{The wavelet transformation is well suited to this problem since it allows us to isolate structures of a given size in an image (in this case, a histogram). \editt{This is in contrast to unsharp masking which only highlights lines and boundaries.}{}} Each velocity plane histogram ($V_R-V_\phi$) is transformed using the Starlet transform \citep{starck94,starck98,starckbook06} and the relative significance of the results are evaluated with respect to Poisson noise \citep{slezak93} and errors in the source \gaia\ data (using Monte Carlo simulations). In this paper, we used a wavelet scale of $8-16$~\kms\ to best identify the Hercules group. See \citetalias{lucchini23} for more details on the wavelet code and methodology.

\begin{figure}
    \centering
    \includegraphics[width=\columnwidth]{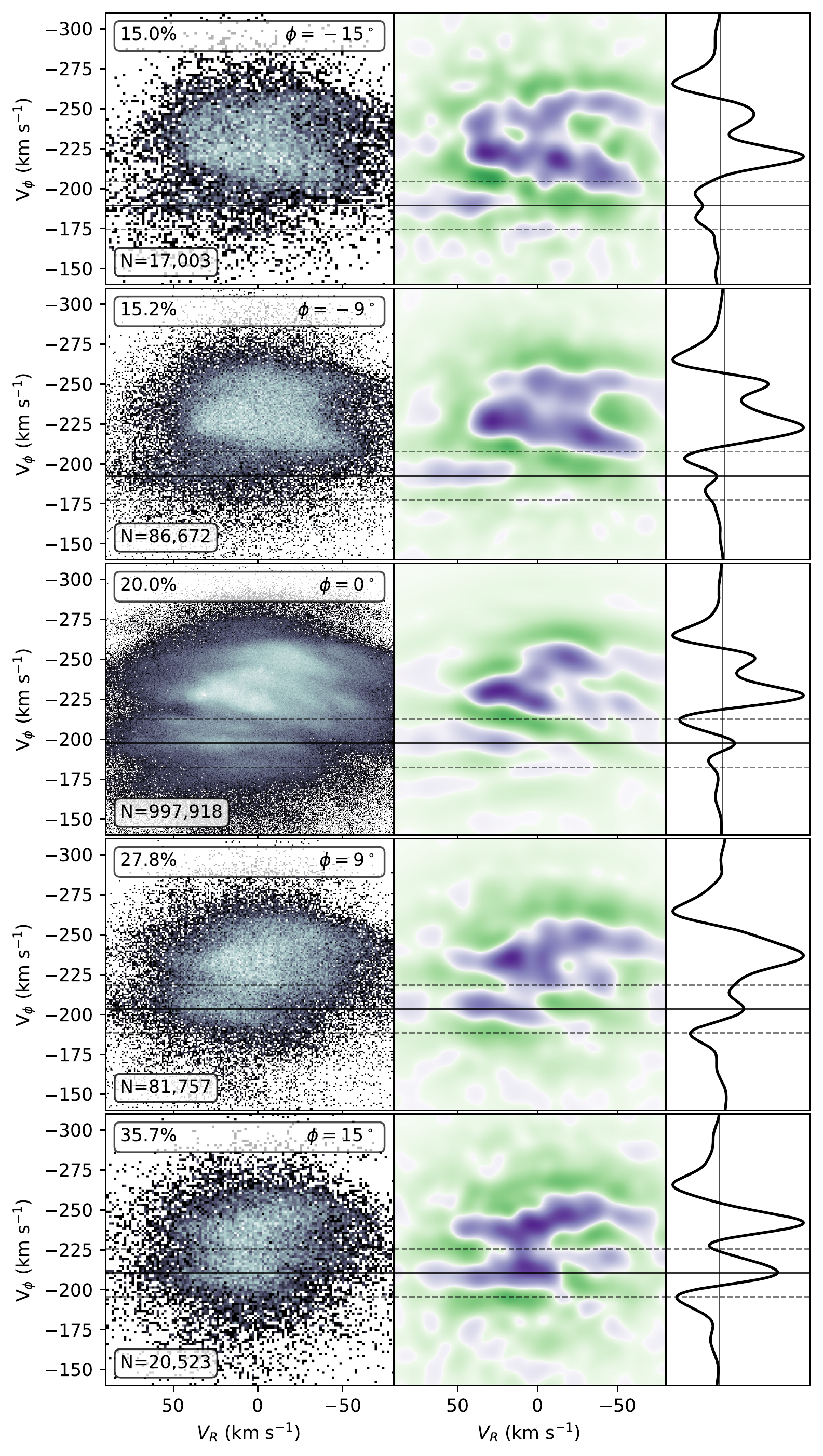}
    \caption{The velocity plane and its wavelet transformation for five neighborhoods located at $\phi=-15^\circ$, $-9^\circ$, $0^\circ$, $9^\circ$, and $15^\circ$. $\phi=0^\circ$ corresponds to the location of the Sun, and \edit{positive}{negative} $\phi$ is in the direction of rotation, while \edit{negative}{positive} $\phi$ is counter to the direction of rotation. The left column shows the $V_R-V_\phi$ histogram for each region, the center column shows the wavelet transformed kinematic plane as in Figure~\ref{fig:results}, and the right column shows the wavelet histogram summed along $V_R$. In the left column, the number of bins in the histogram has been scaled to the total number of stars such that we use 200 bins in each dimension at $\phi=\pm15^\circ$, and we use 600 bins at $\phi=0^\circ$. Hercules is denoted by a horizontal line across each plot in a row, identified as a peak in the summed wavelet histogram (right column). Additionally, the percentage of stars that constitute Hercules (defined as all stars within 15 \kms\ of Hercules' $V_\phi$) is printed in the top left of the left plots. The total number of stars in each bin, N, is also listed in the bottom left of the left plots. \editt{}{This corresponds to mean number densities of 0.2, 1.02, 11.69, 0.96, and 0.24 pc$^{-2}$ in each bin from top to bottom.} In the direction of rotation (i.e. \edit{$+\phi$}{$-\phi$}, towards the major axis of the MW bar\edit{}{, top panels}) Hercules diminishes and the percentage drops, while moving towards the minor axis of the bar (\edit{$-\phi$}{$+\phi$}, towards the Lagrange points\edit{}{, bottom panels}) Hercules grows and merges in with the main mode. We additionally see that $V_\phi$ of Hercules increases as we approach the Lagrange points, as predicted by models (\citealt{donghia20,monari19herc}; see also Figure~\ref{fig:lines}a).}
    \label{fig:phis}
\end{figure}

\section{Results} \label{sec:results}

We track Hercules across the Galactic disk using \gaia\ DR3. In particular, we are looking for variations in the size and strength of the Hercules moving group as we vary the azimuth. Figure~\ref{fig:phis} shows the velocity plane for five different neighborhoods: $\phi=\pm15^\circ$, $\pm9^\circ$, and $0^\circ$ (corresponding to the solar neighborhood shown in Figure~\ref{fig:results}).
These angles were chosen to show the full extent of usable data from \gaia\ DR3 ($\pm15^\circ$, where the number of stars per bin reaches $\sim10^4$), while showing an intermediate region in which Hercules is just beginning to merge with the main mode ($-9^\circ$).
The upper panels are towards the major axis of the bar (\edit{$+\phi$}{$-\phi$}), and the lower panels are towards the minor axis of the bar (\edit{$-\phi$}{$+\phi$}). The left column shows the $V_R-V_\phi$ histogram of all the stars in the specified neighborhood. The center column shows the wavelet transform of these data in which purple represents overdensities and green represents underdensities (as in Figure~\ref{fig:results}). The right-most column shows a plot of wavelet coefficient as a function of $V_\phi$ obtained by summing the wavelet transformed image along $V_R$ (the ``1D summed wavelet histogram''). From these plots, we are able to clearly see the location of Hercules in each neighborhood as the peak in this histogram around $V_\phi\sim-200$ \kms.

\begin{figure}
    \centering
    \includegraphics[width=0.95\columnwidth]{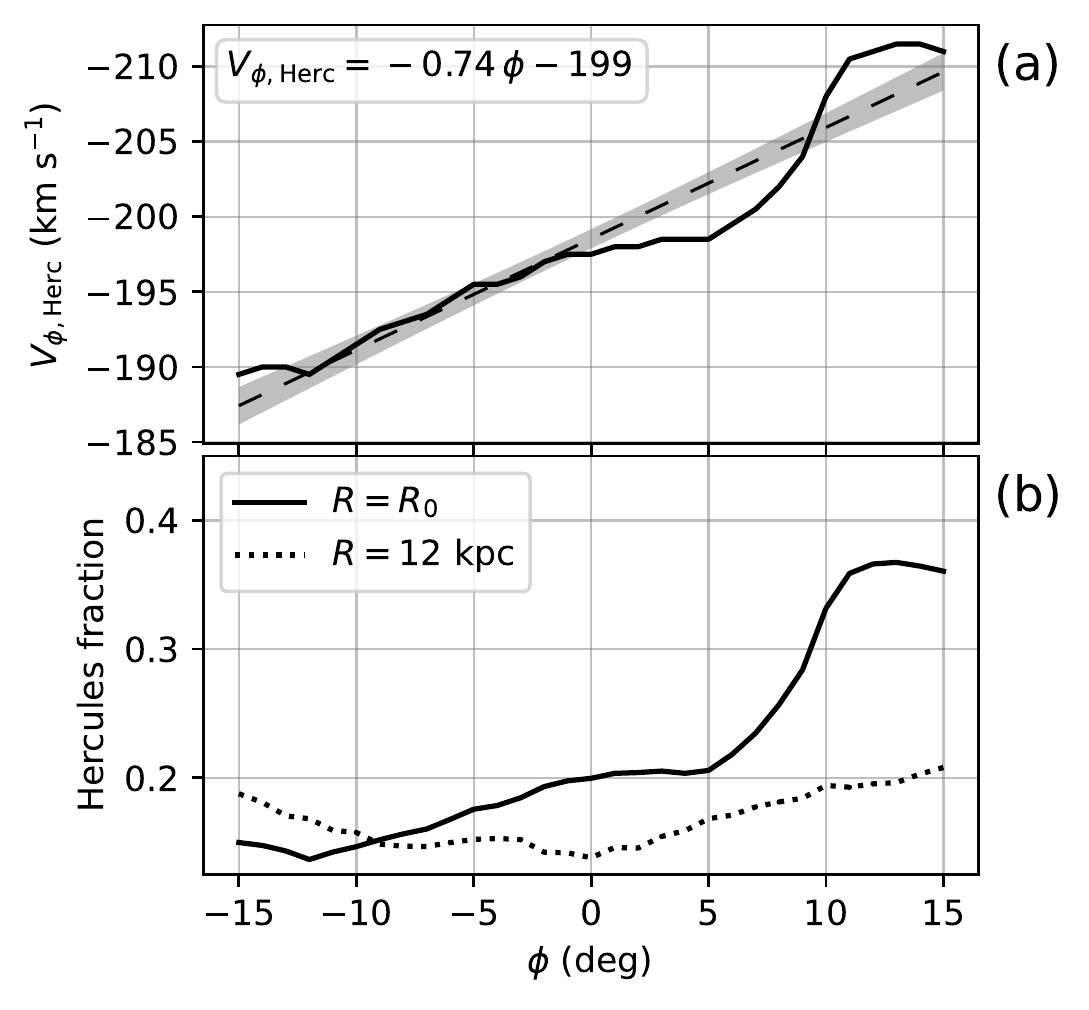}
    \caption{Properties of Hercules as a function of azimuth. The top panel shows the mean $V_\phi$ for the Hercules group as detected using the location of the peak in the wavelet histogram. A linear fit with the 95\% confidence interval is shown as a dashed line and grey shading. The best fit equation is shown in the top \edit{right}{left}. The bottom panel shows the percentage of stars in the given neighborhood that have $V_\phi$ values within 15 \kms\ of Hercules. The dotted line shows the percentage of stars within $175-205$ \kms\ at $R=12$ kpc (such that $R/R_\mathrm{OLR}=0.9$ \editt{}{for a $\sim$40 \kmskpc\ bar}, \editt{as in}{to compare against} \citealt{dehnen00}). At the OLR, we see no variation in intensity as a function of $\phi$. These two plots clearly show that in the $\edit{-}{+}\phi$ direction (towards the bar's minor axis), Hercules sees an increase in angular momentum and becomes more dominant as predicted by the long, slow bar model \citep{donghia20}.}
    \label{fig:lines}
\end{figure}

Figure~\ref{fig:lines} summarizes the properties of Hercules as a function of azimuth.
Figure~\ref{fig:lines}a shows the azimuthal velocity of Hercules as a function of $\phi$. This is identified by the location of the peak in the 1D summed wavelet histogram (right panels in Figure~\ref{fig:phis}). A linear fit to $V_\phi$ of Hercules versus azimuth gives \edit{$V_{\phi\mathrm{, Herc}} = -0.74\,\phi + 198$}{$V_{\phi\mathrm{, Herc}} = -0.74\,\phi - 199$} (with $V_{\phi\mathrm{, Herc}}$ in \kms\ and $\phi$ in degrees) \edit{decreasing from 211 \kms\ at $\phi=-15^\circ$ to 190 \kms\ at $\phi=+15^\circ$}{increasing from 190 \kms\ at $\phi=-15^\circ$ to 211 \kms\ at $\phi=+15^\circ$}.
\editt{Figure~\ref{fig:lines}b shows the strength of the Hercules overdensity relative to the main mode. This is measured by the value of the peak in the 1D summed wavelet histogram corresponding to Hercules. A larger value means that the overdensity is stronger.}{}
In Figure~\ref{fig:lines}\editt{c}{b}, we \editt{also }{}show the percentage of stars that constitute Hercules in each neighborhood. This is determined by selecting all the stars within $\pm15$ \kms\ of the $V_\phi$ of Hercules (Figure~\ref{fig:lines}a) and dividing by the total number of stars in the given bin. \edit{Examples of this 30 \kms\ region}{These 30 \kms\ regions} are shown bounded by the dashed horizontal lines in Figure~\ref{fig:phis}. In the solar neighborhood, Hercules constitutes 20.0\% of stars. As we move towards the bar's major axis (upwards in Figure~\ref{fig:phis}), this percentage decreases to 15.0\% at $\phi=\edit{}{-}15^\circ$. As we move towards the bar's minor axis (downwards in Figure~\ref{fig:phis}), this percentage increases to 35.7\% at $\phi=\edit{-}{+}15^\circ$.

\editt{}{Due to the increase in $V_{\phi\mathrm{, Herc}}$ (Figure~\ref{fig:lines}a), one would naturally expect the fraction of stars in Hercules (within $\pm15$ \kms of $V_{\phi\mathrm{, Herc}}$) to increase since it will get closer to the main mode ($V_\phi\sim-230$ \kms at $R=R_0$). However, by comparing against an unperturbed Dehnen distribution function \citep{dehnen99,hunt18} we see that $V_{\phi\mathrm{, Herc}}$ only increases the Hercules fraction to 25\% compared with 36\% seen in the data. Therefore we conclude that the increase in $V_{\phi\mathrm{, Herc}}$ cannot account for the increase in the Hercules fraction seen in Figure~\ref{fig:lines}b and must be due to the resonance interactions.}

If Hercules was formed through interaction with the Galactic bar's outer Lindblad resonance, we would expect its strength to be relatively constant across azimuth (\citealt{fragkoudi19}, see their Figure~12, left panel \edit{}{showing ``Region 1'' corresponding to Hercules}). However, if the corotation resonance is responsible for Hercules, its member stars would be orbiting around the Galactic bar's L4/L5 Lagrange points \citep{perez-villegas17,donghia20} located along the bar's minor axis. Therefore, we should expect to see a larger \editt{population}{fraction} of stars in Hercules as we approach L4/L5 (in the $\edit{-}{+}\phi$ direction; see Figure~4 from \citealt{donghia20}).
The behavior identified here in \gaia\ DR3, mimics exactly the predictions of \citet{donghia20} (see their Figure~8). \editt{Along}{Towards} the bar's major axis, Hercules diminishes (for both the fast and slow bar scenarios). However, \editt{along}{towards} the minor axis, we are able to discriminate between these two models. For the \editt{short, fast}{high $\Omega_p$} bar, Hercules should remain subdominant and separated from the main mode. While in the \editt{long, slow}{low $\Omega_p$} bar model, Hercules should become extremely prominent, even merging with the main mode, which is what we see in the data.

\editt{Figures~\ref{fig:lines}b and c show}{Figure~\ref{fig:lines}b shows} that Hercules increases in strength \editt{and fraction }{}as we move towards the bar's minor axis ($\edit{-}{+}\phi$). \editt{Figure~\ref{fig:lines}c}{This panel} also shows the fraction of stars within $175-205$ \kms\ at $R=12$ kpc (corresponding to $R/R_\mathrm{OLR}=0.9$ \editt{}{for a $\sim$40 \kmskpc bar}, \editt{as in}{to compare against} \citealt{dehnen00}) to show the angular dependence due to the OLR. As expected, the fraction of stars trapped by the OLR (at $R=12$ kpc) is relatively constant in azimuth, while at the solar radius, there is a sharp increase in the $\edit{-}{+}\phi$ direction, indicating that Hercules is comprised of stars trapped at the Lagrange points.

% \begin{figure}
%     \centering
%     \includegraphics[width=0.8\columnwidth]{figs/Herc_vs_phi_wOLR_190_ci_neg.pdf}
%     \caption{Properties of Hercules as a function of azimuth. The top panel shows the mean $V_\phi$ for the Hercules group as detected using the location of the peak in the wavelet histogram. A linear fit with the 95\% confidence interval is shown as a dashed line and grey shading. The best fit equation is shown in the top \edit{right}{left}. The middle panel shows the relative strength of the Hercules overdensity normalized against the strength of the main mode. This is the value of the peak in the summed 1D wavelet histogram (see right panels in Figure~\ref{fig:phis}). The bottom panel shows the percentage of stars in the given neighborhood that have $V_\phi$ values within 15 \kms of Hercules. The dotted line shows the percentage of stars within 175$-$205 \kms at $R=12$ kpc (such that $R/R_\mathrm{OLR}=0.9$, as in \citealt{dehnen00}). At the OLR, we see no variation in intensity as a function of $\phi$. These three plots clearly show that in the $\edit{-}{+}\phi$ direction (towards the bar's minor axis), Hercules sees an increase in angular momentum and becomes stronger and more dominant as predicted by the long, slow bar model \citep{donghia20}.}
%     \label{fig:lines}
% \end{figure}

Moreover, as shown in \citet{monari19herc}, we would expect the angular momentum of Hercules to vary with azimuth. Figure~\ref{fig:phis} shows that the Hercules overdensity in the data changes its value of $V_\phi$ with angle \edit{decreasing}{increasing} from $V_\phi=\edit{211}{190}$ \kms\ at $\phi=-15^\circ$, to $V_\phi=\edit{190}{211}$ \kms\ at $\phi=+15^\circ$. \citet{monari19herc}\footnote{\citet{monari19herc} give a value of $-8$ \kms\ kpc deg$^{-1}$ for the fit of the angular momentum versus azimuth. In their model the Sun is located at 8.2 kpc giving us a predicted slope of $-8/8.2=-0.96$ \kms\ deg$^{-1}$.} predicted a slope of $-0.96$ \kms\ deg$^{-1}$, whereas here the data show a slope of $-0.74\pm0.04$ \kms\ deg$^{-1}$. This difference could be due to additional effects of spiral arms not included in the model of \citet{monari19herc}. However, it is clear that the slope is $<0$, consistent with Hercules being formed by corotation, not the OLR.

\edit{}{This was also explored in \citet{bernet22} where they found a slope of $-0.5$ \kms\ deg$^{-1}$ for the Hercules group at $R=8$ kpc. Their analysis focused on the detection of arches in kinematic space using a 1D wavelet transformation on data binned along $V_R$. Additionally, their technique to isolate the groups differs from ours which could lead to the slight discrepancy between our result of $-0.74$ \kms\ deg$^{-1}$ and theirs. However, as with the result from \citet{monari19herc}, it is encouraging that we are consistently finding a significant negative slope.}

\editt{}{Figure~\ref{fig:phis} shows that at $\phi=\pm15^\circ$, the number of stars available from \gaia\ is $\sim$2\% that of $\phi=0^\circ$. Therefore the sample of stars will be very different between these regions due to the \gaia\ selection function. Since we are mainly observing the brightest stars at the large distances, the different stellar population observed may affect the properties of Hercules. However, as the fainter stars are usually older (and therefore have had more time to be perturbed by the Galaxy's gravitational potential), we would expect that Hercules would become stronger as we include more faint stars in our sample. This can be tested with \gaia\ DR4 to see how the inclusion of fainter stars at these limits affects Hercules.}

\begin{figure}
    \centering
    \includegraphics[width=1.0\columnwidth]{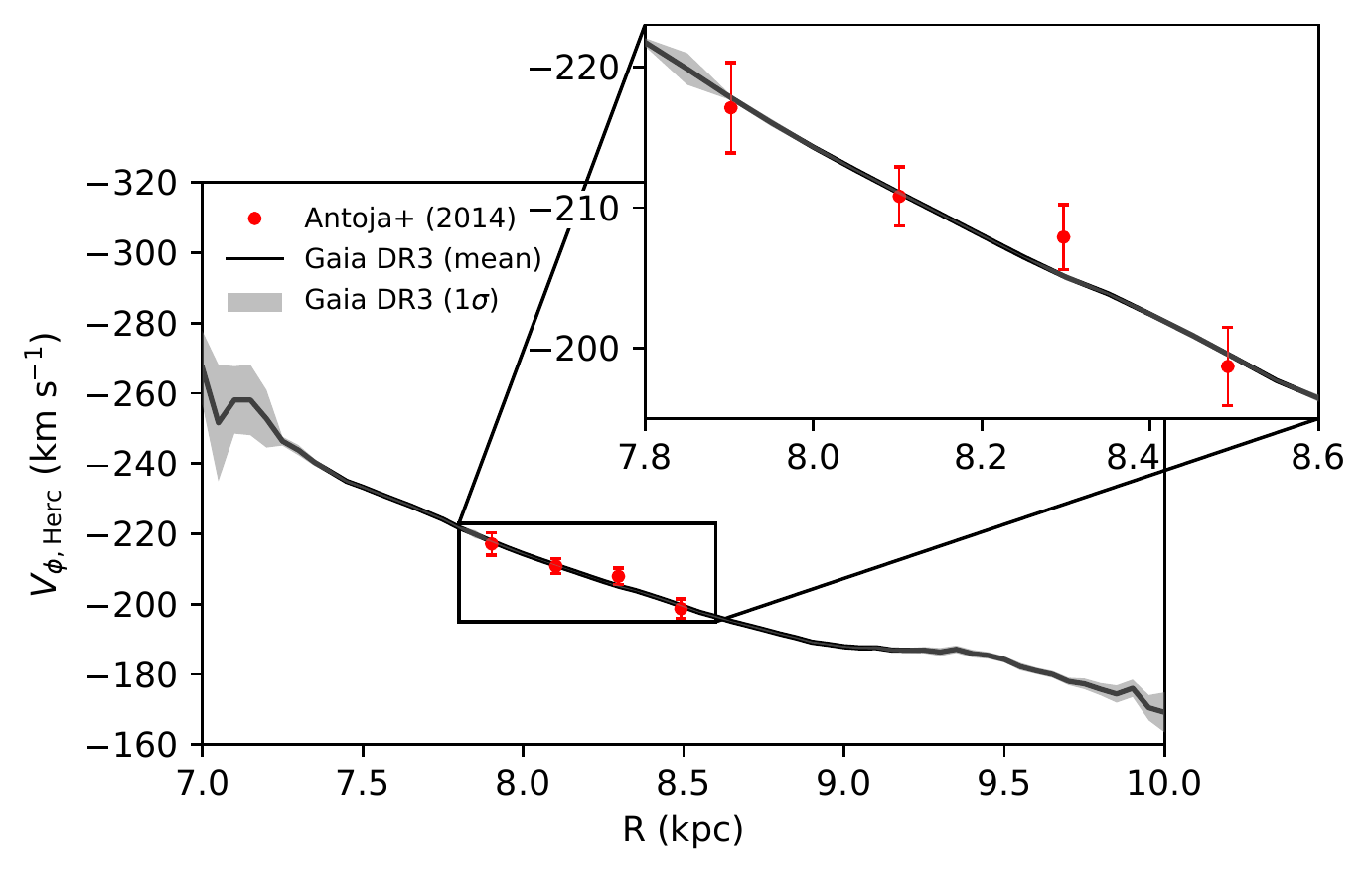}
    \caption{Hercules azimuthal velocity as a function of Galactocentric radius. Following the technique outlined in \citet{antoja14}, we determine $V_\phi$ for the Hercules mode as a function of radius using our \gaia\ DR3 data set at $\phi=6^\circ$. Our results are shown as a solid black line with the grey region marking the 1$\sigma$ error based on 1,000 monte carlo simulations varying the velocities of the stars within their \gaia\ errors. The data points with errors are the measurements from \citet{antoja14}.}
    \label{fig:antoja14}
\end{figure}

\editt{}{While the variation across azimuth is what allows us to discriminate between the low and high $\Omega_p$ bar models, we also briefly explore the change in Hercules' $V_\phi$ as a function of radius to compare with \citet{antoja14}. Figure~\ref{fig:antoja14} shows our calculated $V_\phi$ for Hercules as a function of Galactocentric radius compared against the four data points in \citet{antoja14} from RAVE. We followed the same process as in that paper using $\phi=6^\circ$ (closer to the bar minor axis; see their Figure~8), and determining $V_{\phi,\mathrm{Herc}}$ by finding the minimum of the 1D summed wavelet histogram between the main mode and the Hercules mode. Our measured azimuthal velocities are consistent with those found in \citet{antoja14} (with the largest deviation being 1.2$\sigma$ at 8.3~kpc), however since their exploration exclusively looked at signatures of the OLR from a high $\Omega_p$ bar, this doesn't rule out the low $\Omega_p$ bar model.}

% \begin{figure*}
%     \centering
%     \includegraphics[width=\textwidth]{figs/hercules_angles.pdf}
%     \caption{Three panels showing the evolution of Hercules in azimuth. Each panel shows the wavelet transformed kinematic plane as in Figure~\ref{fig:results} with contours at the $-$1\% (dashed), 5\%, 15\%, 50\%, and 90\% levels. $\phi=0^\circ$ corresponds to the location of the Sun (Panel b), and negative $\phi$ is in the direction of rotation (Panel a) while positive $\phi$ is counter to the direction of rotation (Panel c). In each panel, Hercules is identified, and in the direction of rotation (i.e. towards the major axis of the MW bar) it diminishes, while moving towards the minor axis of the bar (towards the Lagrange points) Hercules grows and merges in with the main mode. We additionally see that $V_\phi$ of Hercules increases as we approach the Lagrange points, as predicted by models \citep{monari19herc}.}
%     \label{fig:phis}
% \end{figure*}

\section{Conclusions} \label{sec:conclusions}

Here we have used \gaia\ DR3 to track the properties of Hercules through Galactic azimuth. Varying Galactocentric $\phi$ 15$^\circ$ either side of the Sun, we see that there is a strong variation in the azimuthal velocity and strength of Hercules. Hercules \editt{becomes stronger and }{}constitutes a larger fraction of stars \editt{per bin}{} as we move towards the minor axis of the bar ($\edit{-}{+}\phi$). This is in direct agreement with predictions of a \editt{long, slow}{low pattern speed} bar model in which Hercules is formed through stars trapped at corotation, orbiting the L4/L5 Lagrange points. This \edit{}{independently} corroborates \edit{recent direct observations of the bar from \gaia\ DR3}{several recent works converging on the slow bar model \citep{monari19herc,binney20,chiba21,chiba21b,drimmel22}. Future analytic work exploring the effect of the bar on Hercules across a full parameter space of models will also give us more direct constraints on its properties.}

With the next release from \gaia, DR4, we can expect further significant improvements using this technique. While DR3 extended our view out to $\phi=\pm15^\circ$, we can hope to get similar signal to noise results out to $\phi=\pm30^\circ$ and beyond. This will allow us to get a direct measurement of the bar angle with respect to the Sun's position, by looking for a minimum in Hercules' strength and $V_\phi$.

\section*{Acknowledgements}
The authors thank the anonymous referee for their constructive comments on the manuscript. J.A.L.A. has been founded by The Spanish Ministerio de Ciencia e Innovacion by the project PID2020-119342GB-I00.
This work made use of Astropy:\footnote{http://www.astropy.org} a community-developed core Python package and an ecosystem of tools and resources for astronomy \citep{astropy:2013, astropy:2018, astropy:2022}.

\section*{Data Availability}
The data underlying this article will be shared on reasonable request to the corresponding author.

\nolinenumbers
% \newpage
\bibliographystyle{mnras}
\bibliography{references}

\label{lastpage}

\end{document}